\documentclass[letterpaper,usenatbib]{mn2e}

\usepackage{graphicx}
\newcommand{\HI}{H{\,\small I}}
\newcommand{\HII}{H{\,\small II}}
\newcommand{\OI}{O{\,\small I}}
\newcommand{\OII}{O{\,\small II}}
\newcommand{\OIII}{O{\,\small III}}
\newcommand{\SII}{S{\,\small II}}
\newcommand{\NII}{N{\,\small II}}
\newcommand{\Hb}{H$\beta$}
\newcommand{\Ha}{H$\alpha$}
\title[jet-induced outflow in 3C~293]{A jet-induced outflow of warm gas in 3C~293}
\author[B.H.C.Emonts et al.]{B.H.C.Emonts$^{1}$\thanks{E-mail:
emonts@astro.rug.nl}, R.Morganti$^{2}$, C.N.Tadhunter$^{3}$, T.A.Oosterloo$^{2}$, 
\newauthor J.Holt$^{3}$ and J.M.van der Hulst$^{1}$\\
$^{1}$Kapteyn Astronomical Institute, University of Groningen, P.O. Box 800, 9700 AV Groningen, The Netherlands\\
$^{2}$Netherlands Foundation for Research in Astronomy, Postbus 2, 7990 AA Dwingeloo, The Netherlands\\
$^{3}$Department of Physics and Astronomy, University of Sheffield, Sheffield S3 7RH, UK}
\begin{document}

\date{}

\pagerange{\pageref{firstpage}--\pageref{lastpage}} \pubyear{2004}

\maketitle

\label{firstpage}

\begin{abstract}
Using long slit emission-line spectra we detect a fast outflow of ionized gas, 
with velocities up to 1000 km s$^{-1}$, in the nearby powerful radio galaxy 3C~293 ($z=0.045$). 
The fast outflow is located about 1 kpc east of the nucleus, in a region of enhanced radio 
emission due to the presence of a distorted radio jet. We present results that indicate that this 
fast outflow is caused by a jet-ISM interaction. The kinematics of the outflowing ionized gas are 
very similar to those of a fast outflow of neutral hydrogen gas in this galaxy, suggesting 
that both outflows are the result of the same driving mechanism. While the mass of the 
outflowing ionized gas is about $1 \times 10^{5}$ $M_{\odot}$, the total \HI\ mass involved in the 
neutral outflow is about $100 \times$ higher (10$^{7}$ $M_{\odot}$). This shows that, despite the high 
energies that must be involved in driving the outflow, most of the gas remains, or becomes 
again, neutral. Other outflows of ionized gas, although not as pronounced as in the region of the 
enhanced radio emission, are also seen in various other regions along the axis of the inner radio 
jets. The regular kinematics of the emission-line gas along the major axis of the host galaxy 
reveal a rotating ionized gas disk 30 kpc in extent.
\end{abstract}

\begin{keywords}
galaxies: active - galaxies: individual: 3C~293 - galaxies: ISM - ISM: jets and outflows - ISM: kinematics and dynamics - line: profiles
\end{keywords}

\section{INTRODUCTION}
\label{sec:intro}

Nuclear activity can strongly influence the Interstellar Medium (ISM) in the centre 
of active galaxies. The strong radiation field from the central engine, as well as 
collimated jets of radio plasma, affect the ionization and kinematics of this ISM in a 
complex way. In addition, the situation can be further complicated by the presence 
of starburst-induced winds \citep[e.g.][and references therein]{hec90}, associated with young stellar populations that are known to be present in some radio galaxies \citep[e.g.][]{are01,wil02,wil04}. The feedback mechanisms of AGN and starburst activity are also important in the evolution of the host galaxies. AGN activity may, for example, regulate the correlation between the mass of the central black hole and the galaxy's bulge properties \citep[e.g.][]{sil98}. It is therefore important to carefully study these physical processes and determine to what extent each of them influences the characteristics of the host galaxies. Nearby radio galaxies provide excellent opportunities to do this in detail. 

An important feature that is observed in a significant fraction of active galaxies is the presence of outflows of warm and hot gas from the central region \citep[e.g.][and references therein]{hec81,vei02,kri04}. In a number of nearby powerful radio galaxies asymmetric emission-line profiles provide evidence for outflows of ionized gas, often with velocities $>10^{3}$ km~s$^{-1}$ \citep[e.g.][]{tad91,vil99,tad01,hol03,tay03}. Recently, a number of nearby radio galaxies have been found to contain such fast outflows not only in the ionized gas, but also in {\sl neutral} hydrogen (\HI) gas. Among these radio galaxies are 4C~12.50 \citep[PKS 1345+12;][]{mor04a} and 3C~305 \citep{mor05}. An overview of the \HI\ outflows found so far in radio galaxies is given by \citet{mor04c}. The exact driving mechanisms for these fast outflows of neutral and ionized gas are not always clear, but they could consist of interactions between the ISM and the propagating radio plasma \citep[like the case of Seyfert galaxy IC 5063;][]{oos00,mor04b}, AGN induced winds \citep[e.g.][]{kro86,bal93,dop02} or starburst-related phenomena \citep[e.g.][]{hec90}. The detection of neutral gas involved in the fast outflows puts serious constraints on the, most likely very energetic, driving mechanisms.

An excellent case to study the physical mechanisms in detail is the nearby powerful radio galaxy 3C~293, in which we recently detected an outflow of neutral hydrogen gas from the central region \citep[][hereafter Paper
1]{mor03}. Velocities of the outflowing \HI\ gas reach $\sim 1000$ km
s$^{-1}$, making it one of the most extreme cases of outflowing \HI\ presently
known. Unfortunately, due to limited spatial resolution of the observations we
were not able to determine the exact location of this \HI\ outflow. In this
paper we analyse the characteristics of the ionized gas in 3C~293 and relate its kinematics with that of the outflowing neutral gas. We aim to gain a better understanding of the physical process that is responsible for the outflow and the effect it has on the ISM of the host galaxy. 

3C~293 is a powerful ($P_{\rm 1.4GHz}\sim2\times10^{25}$ W~Hz$^{-1}$)\footnote[1]{$H_{\circ} =
71$ km~s$^{-1}$~Mpc$^{-1}$ used throughout this paper. At the redshift of
3C293, $z = 0.045$, this puts the galaxy at a distance of 190 Mpc and 1 arcsec
= 0.92 kpc.}, edge-brightened Fanaroff-Riley type-{\,\small II} \citep{fan74} radio source with two extended  radio lobes \citep[e.g.][see
also Figure \ref{fig:slitposition}]{bri81}. It also has a Steep Spectrum Core that consists of a true flat-spectrum radio core and a two-sided radio jet structure which is distorted and misaligned with respect to the outer radio lobes \citep[e.g.][see also Figure \ref{fig:slitposition}]{aku96,bes02}. 3C~293 is hosted by the early type galaxy UGC 8782, that has a
complex disk-like morphology with a bridge-tail structure that extends beyond
a possible companion $\sim 33$ arcsec to the southwest
\citep{bre84,hec86,eva99}. The spectroscopic work of \citet{tad05} reveals
post-starburst young stellar populations with ages between 0.1 and 2.5 Gyr
throughout the galaxy. 3C~293 has a modest far-IR luminosity of
$L_{\rm fir} \sim 2.3 \times 10^{10}$ $L_{\odot}$. The galaxy contains
extensive filamentary dust lanes \citep[e.g.][]{mar99} and large amounts of molecular and neutral hydrogen gas in the central few kpc \citep{baa81,has85,eva99,bes02,bes04}. Given its peculiar optical morphology
and the presence of large amounts of cold gas and young stars, 3C~293 has
often been suggested to have been involved in a gas-rich galaxy-galaxy
interaction or merger event, in which large amounts of gas have been funnelled
into the nuclear region and are now fuelling the AGN. The presence of a dense
ISM, a young stellar population and a distorted radio morphology make 3C~293
an excellent case to study the interplay between ISM, nuclear starburst and
AGN activity.

\section{OBSERVATIONS}
\label{sec:observations}

\begin{figure}
\includegraphics[width=8cm]{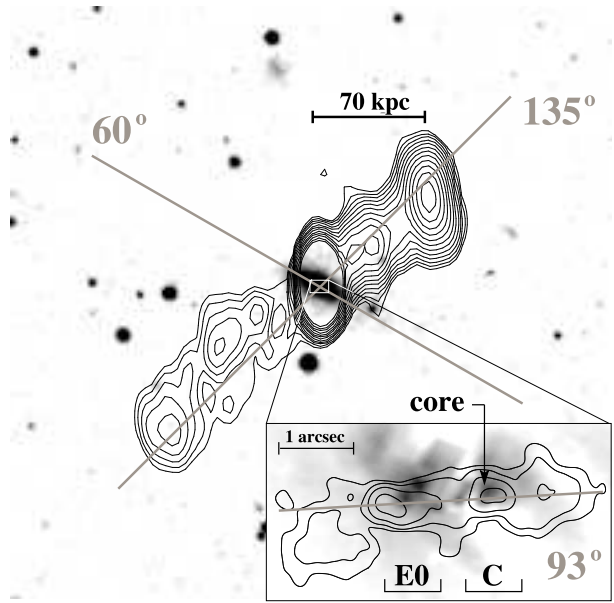}
\caption{Position of the slits along 60$^{\circ}$ (major axis host galaxy), 135$^{\circ}$ (outer radio axis) and 93$^{\circ}$ (inner radio axis). The large-scale image is an overlay of WSRT radio contours (constructed from our 10 MHz data) superposed onto an optical DSS image of 3C~293. Contour levels in the WSRT image are 4.5, 6.5, 9.0, 13, 18, 25, 35, 50, 70, 95, 130, 170 mJy~beam$^{-1}$. The zoom-in on the central region is an overlay of a MERLIN radio continuum image onto a HST F702W broadband image \citep[overlay from][see Section \ref{sec:radioaxis} for more details]{bes02}. Contour levels in the MERLIN image are 9, 36, 144, 630 mJy beam$^{-1}$. The regions C and E0 are the region of the core and the region of bright optical and radio emission 1 arcsec east of the core. These regions are described in detail in Section \ref{sec:radioaxis}.}
\label{fig:slitposition}
\end{figure}
\begin{table}
\caption{Observational Parameters}
\label{tab:obsparam}
\begin{tabular}{cccccc}
 & & slit & integration & &\\
date &  p.a. & width & time (s)  &  average & seeing\\
 &  ($^{\circ}$) & ('') & blue $\ $ red &airmass & ('') \\
\hline   
5/12/2001 & 60  & 1.3 & 2400 $\ \ $ 2400 & 1.085 & $\sim$ 0.8 \\
5/12/2001 & 135 & 1.3 & 1200 $\ \ $ 1200 & 1.037 & $\sim$ 0.8 \\
6/18/2003 & 93  & 1.03& 3600 $\ \ $ 2400 & 1.02  & $\sim$ 1.0 \\          
\end{tabular} 
\end{table} 

Optical long-slit spectra were taken at the William Herschel Telescope
(WHT\footnote[2]{The WHT is operated on the island of La Palma by the Isaac
Newton Group in the Spanish Observatorio del Roque de los Muchachos of the
Instituto de Astrofisica de Canarias.}) on May 12th 2001 (p.a. 60$^{\circ}$
and 135$^{\circ}$) and in service mode on June 18th 2003 (p.a. 93$^{\circ}$)
using the ISIS long-slit spectrograph with the 6100\AA\ dichroic, the R300B
and R316R gratings at the blue and red arm and the GG495
blocking filter on the red arm to cut out second order blue light. This
resulted in a wavelength coverage from about 3500 to 8000 \AA. The slit was aligned by eye and centred on the brightest part of the
galaxy. The different slit positions are indicated in Figure
\ref{fig:slitposition}. Table \ref{tab:obsparam} gives a summary of the
observational parameters.

We used the Image Reduction and Analysis Facility (IRAF) to reduce the data. A
description of the data reduction of the May 12th 2001 data is already given
by \citet{tad05}. For the June 18th 2003 data (p.a. 93$^{\circ}$) the standard
calibration (bias subtraction, flatfielding and wavelength calibration) was
done. We used dome and arc exposures taken at approximately the same position
and time as the source spectra. We took out a significant tilt of the slit in the spatial direction by applying a 1st order correction using two stars that were in the slit. With the tilt removed, the spectra are aligned within one pixel. After the background subtraction the frames were combined and cosmic rays removed. Two standard stars (BD+26 2606 and BD+33 2642) were used for the flux calibration. The wavelength calibration was checked for both the 2003 and 2001 data using night skylines in spectra that were reduced without background subtraction. We corrected for any systematic off-set of the lines in the different spectra (likely due to flexure in the spectrograph) during the analysis of the spectra. The resulting accuracy of the wavelength calibration is within 0.1~\AA\ for all the red spectra and 0.4~\AA\ for all the blue spectra, with no obvious systematic errors. The $\lambda$-resolution of the spectra is $\sim 4$\AA. 

For the analysis of the spectra, we used the Starlink package FIGARO to bin
consecutive rows of pixels into a one-dimensional spectrum. In this way, we
produced a series of spectra across the slit. These spectra were further
analysed with the Starlink package DIPSO by fitting Gaussian profiles to the
spectral lines. Velocities used in this paper are heliocentric velocities.

In order to study neutral hydrogen gas in emission around 3C~293 we obtained
Westerbork Synthesis Radio Telescope (WSRT\footnote{The WSRT is operated by
the Netherlands Foundation for Research in Astronomy with support from the
Netherlands Foundation for Scientific Research.}) observations on May 7th
2001, using the 10 MHz band. The goal of this study was to look for possible \HI\ structures that could be
reminiscent of a merger event in 3C~293. In the line data we detect three
nearby companions of 3C~293 in \HI\ emission, which we describe in detail in
\citet{emo04}. However, as explained in \citet{emo04}, the bandwidth of these
observations turned out to be not ideal for looking for \HI\ emission directly
associated with the host galaxy of 3C~293. In the present paper we use the
WSRT data to make a new continuum image of 3C~293. The channels containing any
deep \HI\ absorption were excluded from the continuum fitting. We used the
MIRIAD software to make the continuum image shown in Figure
\ref{fig:slitposition}, which we use as a reference to show the slit
alignments. The detection level of the radio continuum is limited by a dynamic
range $\sim 500$, while the peak intensity is 3.6 Jy~beam$^{-1}$. The beam-size
is $25.3 \times 11.9$ arcsec at p.a. 1.1$^{\circ}$. The total power is $P_{\rm 1.4GHz} \approx 2 \times 10^{25}$ W~Hz$^{-1}$.

\section{RESULTS} 
\label{sec:results}

\begin{figure}
\includegraphics[width=8cm]{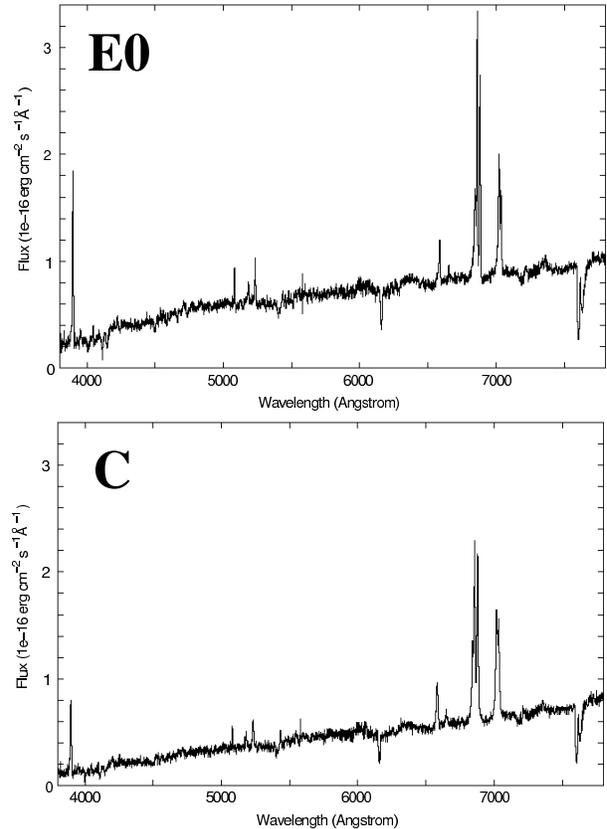}
\caption{Spectrum of the regions E0 and C. The spectrum is a composite spectrum of both the red and the blue spectrum along p.a. 93$^{\circ}$. The flux is summed over an aperture of 0.8 arcsec. See Section \ref{sec:ionization} for more details on these spectra.}
\label{fig:specartfin2}
\end{figure}
Extended continuum and emission-lines are detected in all our optical
spectra of 3C~293. Figure \ref{fig:specartfin2} shows a typical spectrum of 3C~293 in the regions E0 and C (these regions are indicated in Figure \ref{fig:slitposition} and will be described in detail below). In this Section various features of the emission-line gas in 3C~293 along the different slits are presented.

For the analysis of the kinematics of the gas we will mainly use the [\SII] and [\OII] lines. These are among the stronger emission lines detected and they do not suffer from absorption features in the underlying continuum as can be the case for the stronger H$\alpha$ line.

\subsection{Gas kinematics along the inner radio axis}
\label{sec:radioaxis}

\begin{figure}
\includegraphics[width=8cm]{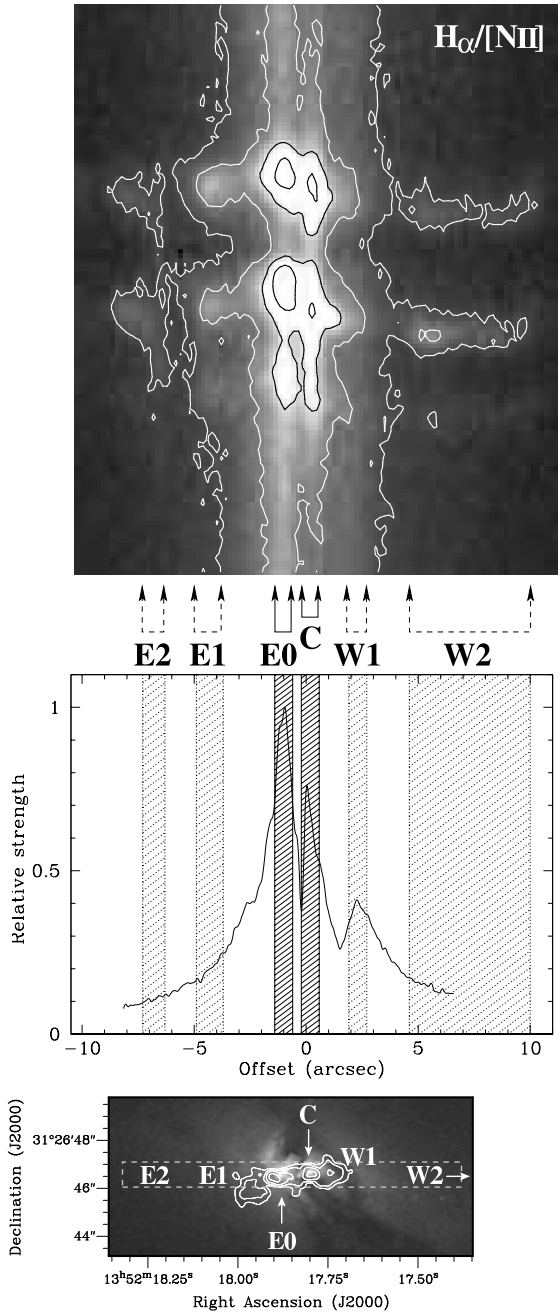}
\caption{Bottom: MERLIN radio contour map overlaid on an optical HST image of the inner few kpc of 3C~293 \citep[figure from][]{bes02}. Contour levels are 9, 36, 144, 630 mJy~beam$^{-1}$. Middle: Integrated profile from the HST image across the region that is covered by the 93$^{\circ}$ slit (indicated by the dotted region in the bottom plot). Top: Spectrum along p.a. 93$^{\circ}$ (H$\alpha$+[\NII]). Contour levels: 0.75, 1.67, 3.4, 6.1 $\times$ 10$^{-17}$ erg$\ $cm$^{-2}$s$^{-1}$\AA$^{-1}$. {\sl The features in the spectrum can be traced in the HST image.}}
\label{fig:overlay}
\end{figure}

In this section we concentrate on the analysis of the kinematics of the
gas along the inner radio axis, i.e. p.a. 93$^{\circ}$. Along this position
angle the ionized gas clearly shows the most complex kinematics.  This is
illustrated by the 2D spectrum of the H$\alpha$+[\NII]$\lambda$$\lambda$6548,6583 region shown in Figure
\ref{fig:overlay}.  In this figure, the locations of the various features
detected in the optical spectrum are compared with features seen in the radio and the HST image.
Figure \ref{fig:overlay} (bottom) shows a MERLIN radio contour map overlaid on
an optical HST image of the inner 5 kpc of 3C~293 \citep[figure from][also
shown in Figure \ref{fig:slitposition}]{bes02}. The accuracy of the overlay is
within 0.3 arcsec. The HST WFPC2 image was taken with a broadband F702W filter
and includes continuum in the wavelength range just below the strong
H$\alpha$+[\NII] lines. In Figure \ref{fig:overlay}
(middle) an integrated brightness profile from the HST image is taken across the region
that is covered by the 93$^{\circ}$ slit, as indicated by the dotted region in
the HST image. Figure \ref{fig:overlay} (top) shows that the different
features in the HST image can be identified in the red part of our 2D-spectrum
along p.a. 93$^{\circ}$.  Several regions of interest are
indicated. Region C, when observed in the radio at VLBI resolution
\citep{bes04}, includes the actual core as well as a radio knot about 0.4
arcsec west of the nucleus. Region E0 is the region of peak intensity in the
optical, about 1 arcsec east of the nucleus. This is also the region
in which the radio continuum peaks. Regions E1 and W1 are regions east and west of
the nucleus just at the edge of where fainter radio continuum is detected. Further out to the east and west (in regions where so far no radio
continuum has been detected) are the regions E2 and W2.

In Figure \ref{fig:SIIfit} we show the 2D-spectrum of the [\SII] doublet. Also shown in Figure \ref{fig:SIIfit} are fits to the [\SII] line at various places along the slit. As mention above, the [\SII] line is used for the kinematic analysis in order to avoid possible effects of absorption features that may affect the H$\alpha$ line. Moreover, the region of the nucleus is more clearly seen in the red part of the spectrum than in the blue part \citep[due to the higher sensitivity of the ISIS spectrograph in the red as well as complex extinction across the spectral range due to prominent central dust-lanes;][]{mar99}, which favours the use of [\SII] above [\OII] for the detailed kinematic analysis presented in this Section. 

\begin{figure*}
\includegraphics[width=18cm]{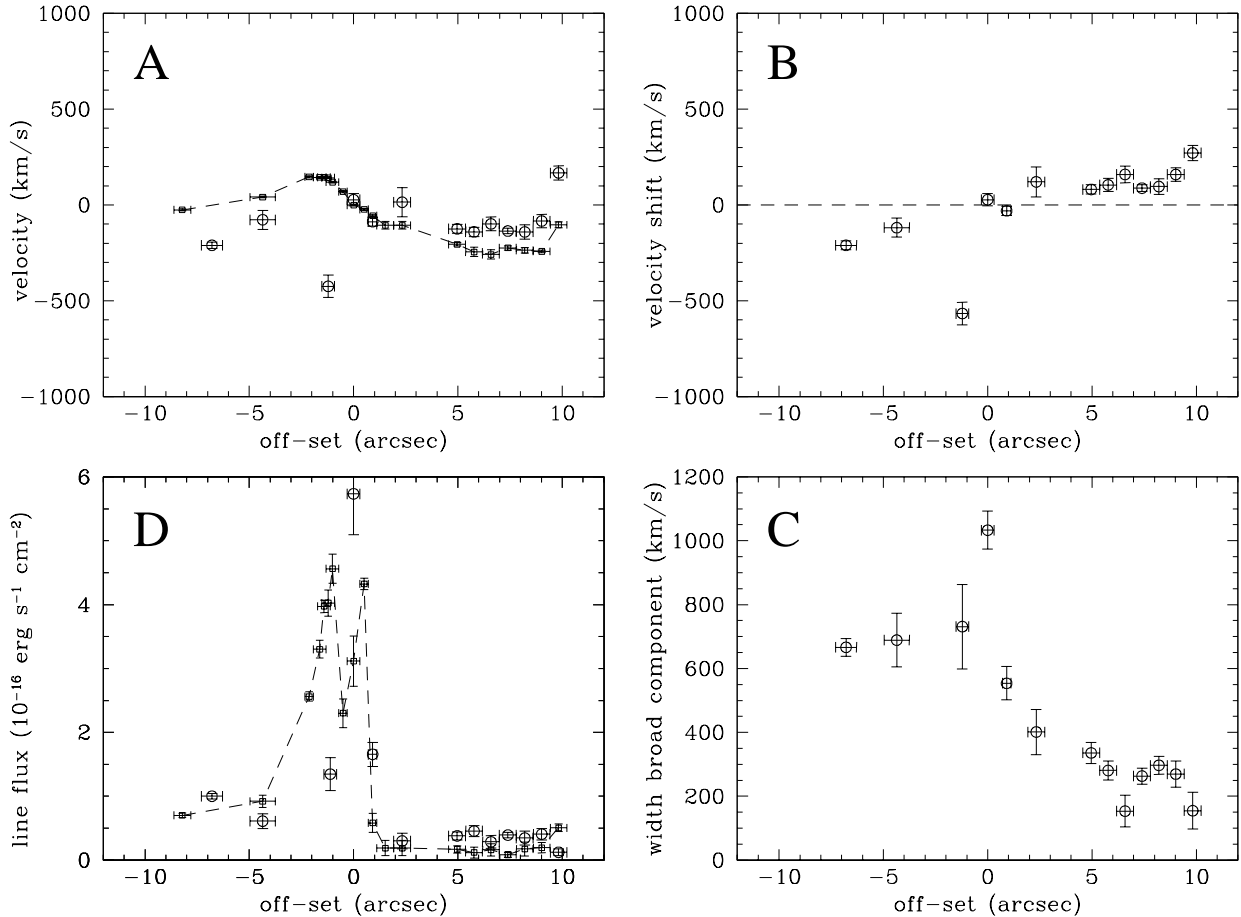}
\caption{Line-fitting results of the [\SII] line in the spectrum along 93$^{\circ}$. In the various plots the small squares (connected with a dotted line) represent the narrow component of the gas. The circles represent the broad component. Vertical error bars represent uncertainties in the fits and in the $\lambda$-calibration; horizontal error bars represent the width of the used apertures. The horizontal axis is centred on the nucleus. Plotted against the spatial off-set from the nucleus are: (a) the velocity of both the narrow and the broad component w.r.t. the systemic velocity ($v_{\rm sys} = 13,450 \pm 35$ km~s$^{-1}$; see Section \ref{sec:radioaxis}); (b) the velocity shift between the broad and narrow component (the dotted line, velocity shift = 0, corresponds to the velocity of the narrow components); (c) the width of the broad component; (d) the [\SII]$\lambda$6716+$\lambda$6731 line flux (mean value over an aperture of 0.2 arcsec) of both the narrow and broad component.}
\label{fig:artSII93b}
\end{figure*}

In Figure \ref{fig:SIIfit} we can immediately identify a narrow component in most of the fits. To fit the narrow component of the [\SII] doublet we use two
Gaussians with equal width and with wavelength separation constrained,
following the physical parameters (corrected for the redshift of 3C~293) given
by \citet{ost89}. At most places along the slit we need a second, broader component to get a good fit (again this is a doublet component with equal width and wavelength separation constrained). Figure \ref{fig:artSII93b} shows various results from the line-fitting of both components of the [\SII] line in the spectrum along 93$^{\circ}$. Plotted against the distance of the features from the nucleus are the velocities of both the narrow and broad component relative to the systemic velocity of 3C~293, the velocity shift between the narrow and broad component, and the width of the broad component along the slit. Also shown is the line flux of both the narrow and broad components. Table \ref{tab:valuescomp} summarizes the kinematic information of both components in the different regions.

From Figures \ref{fig:SIIfit} and \ref{fig:artSII93b} and Table \ref{tab:valuescomp} we can distinguish several features with distinct gas kinematics. In Figure \ref{fig:artSII93b}(a), the narrow components (connected with the dashed line) trace what appears to be a regular rotating feature that we identify with the large scale gas disk already observed by \citet{bre84}. We will discuss more about this disk in Section \ref{sec:disk}. In the following we briefly discuss the characteristics of the ionized gas in each region:

\begin{figure*}
\includegraphics[width=17.5cm]{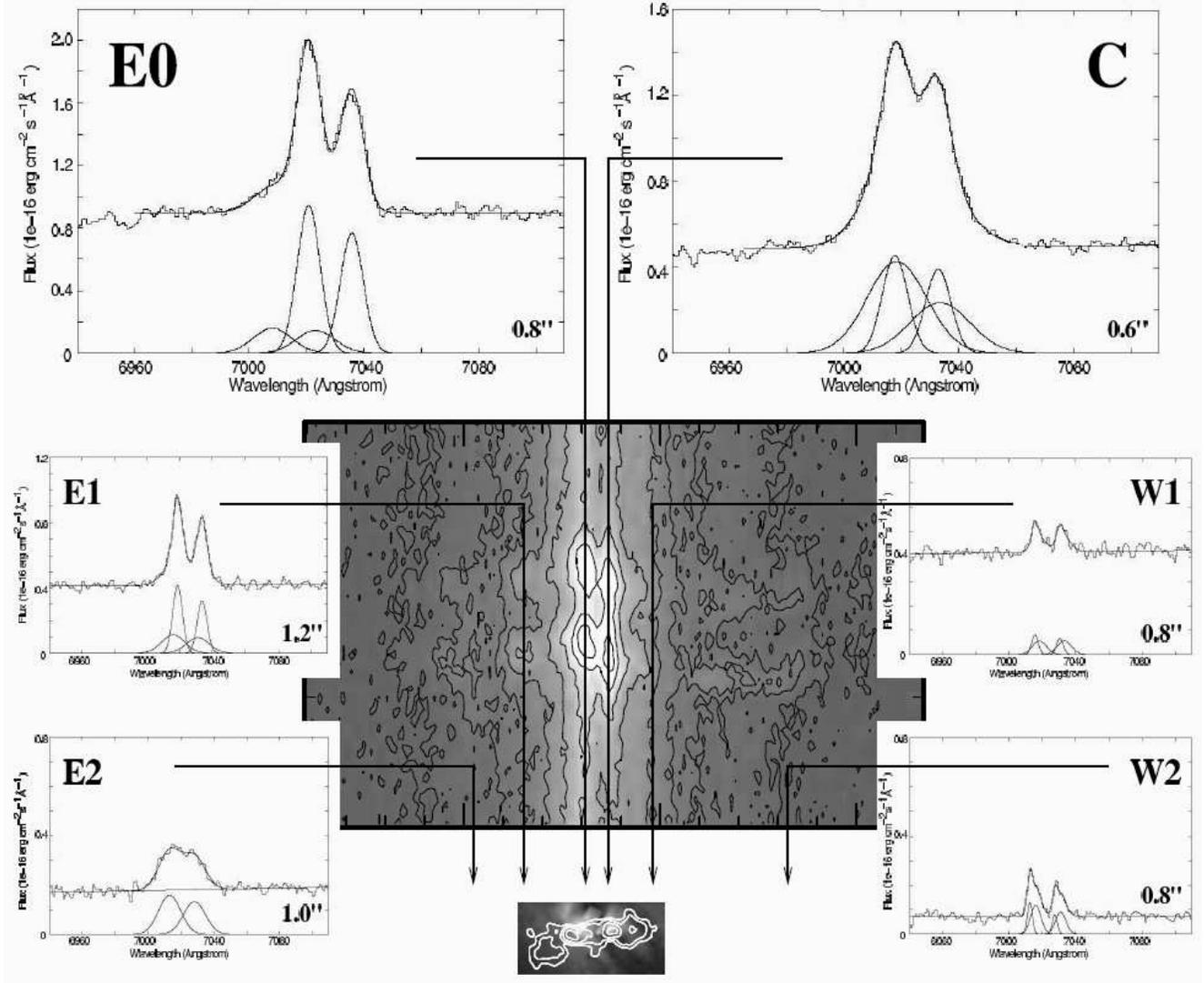}
\caption{Spectrum of the [\SII]$\lambda$$\lambda$6716,6731 line (middle; contour levels 0.2, 0.48, 0.72, 1.1, 1.6, 2.5, 3.6, 4.6 $\times$ 10$^{-17}$ erg$\ $cm$^{-2}$s$^{-1}$\AA$^{-1}$). Also shown are the fits to the [\SII] line at various places along the slit. At most places along the slit we need two components to get a good fit. The apertures used to make the 1D-spectra that we present here are written in the bottom-right corner of the plots (note that the plot of region W2 is taken with an aperture of 0.8 arcsec, but a similar spectrum is seen throughout the region W2 as it is indicated in Figure \ref{fig:overlay}).}
\label{fig:SIIfit}
\end{figure*}
\begin{table*}
\caption{Kinematic information of the various narrow and broad components observed in the different regions. Given is the FWHM (corrected for instrumental broadening) of the fit to both components as well as the velocity shift ($\Delta v$) between the centre of the broad component and the centre of the narrow component in each region (negative shift means that the broad component is blueshifted). Also given is the [\SII]$\lambda$6716+$\lambda$6731 flux of the various components (given as the calculated mean flux over an aperture of 2 arcsec). Due to the faintness of the narrow component across region W2 we could not obtain a single good value for the width of this narrow component, therefore we left it out of this Table. {\sl Note that the broad component represents different gas features in the different regions (see text).}}
\label{tab:valuescomp}
\begin{tabular}{c|c|c|c|c|c}
Region & FWHM$_{narrow}$ & FWHM$_{broad}$ & $\Delta v$ & Flux$_{narrow}$ & Flux$_{broad}$ \\ 
       & (km s$^{-1}$)  & (km s$^{-1}$)   & (km s$^{-1}$) & ($\times$10$^{-16}$ erg$\ $s$^{-1}$$\ $cm$^{-2}$) & ($\times$10$^{-16}$ erg$\ $s$^{-1}$$\ $cm$^{-2}$) \\
\hline   
E2 & - & 664 $\pm$ 27 & -202 $\pm$ 22 & - & 1.00 $\pm$ 0.04 \\
E1 & 248 $\pm$ 15 & 687 $\pm$ 80  & -114 $\pm$ 48 & 0.90 $\pm$ 0.10 & 0.61 $\pm$ 0.12 \\ 
E0 & 341 $\pm$ 10 & 729 $\pm$ 127 & -542 $\pm$ 56 & 4.02 $\pm$ 0.20 &  1.35 $\pm$ 0.26 \\
C  & 411 $\pm$ 20 & 1032$\pm$ 57  & 26 $\pm$ 31 & 3.12 $\pm$ 0.39  & 5.73 $\pm$ 0.64 \\
W1 & 112 $\pm$ 55 & 398 $\pm$ 68 & 115 $\pm$ 75 & 0.19 $\pm$ 0.12 & 0.30 $\pm$ 0.12 \\ 
W2 &  - & 264 $\pm$ 40 & 152 $\pm$ 33 & - & 0.41 $\pm$ 0.09 \\
\end{tabular}
\end{table*} 
  
\begin{itemize}
  
\item{{\bf Region C:} Region C is the region that includes the nucleus. In region C we need to
    include a broad component to get a good fit to the emission lines
    (Figure \ref{fig:SIIfit}). This broad component, however, is
    strong compared to the broad component at other places along the
    slit and it peaks at the same velocity as the narrow
    component. It therefore might reflect a more turbulent
    state of the ionized gas close to the nucleus of the galaxy or the
    presence of a circum-nuclear disk. Similar phenomena have been suggested to to explain a broadening of the deep \HI\ absorption profile in the nuclear region of 3C~293 \citep{bes02}.
  
At the location of the nucleus, we very accurately determine the
  systemic velocity of 3C~293 to be $v_{\rm sys} = 13,450 \pm 35$ km~s$^{-1}$ or $z =
  0.04486 \pm 0.00012$, which is the redshift of the narrow component in region C. The error has been determined from the uncertainty in
  fitting the Gaussian components, from the inaccuracy of the wavelength
  calibration and from the uncertainty in the exact position of the nucleus in the optical spectrum.}

\item{{\bf Region E0:} Region E0 is the region about 1 kpc east of the nucleus, where both the optical as well as the radio continuum peak in intensity. In region E0 we detect, in addition
to a narrow component, a broad component of the [\SII] line that is highly
{\sl blueshifted} with respect to the velocity of the narrow component and the systemic velocity of the system. A similar component is seen in the [\OII] lines. The FWHM (full-width at half the maximum intensity) of this
component is $729 \pm 127$ km~s$^{-1}$ (corrected for instrumental broadening) and its centre is blueshifted by $542
\pm 56$ km~s$^{-1}$ w.r.t. the velocity of the narrow component
that traces the regular rotating gas disk (see Figure
\ref{fig:artSII93b}).  
The width of the broad component and the amplitude of the velocity shift (w.r.t the narrow component) are so large that they cannot be due to gravitational motions of
the gas. Thus, we interpret this broad, blueshifted component as a fast
outflow of ionized gas. The exact extent of this outflow - both along
the radio jet as well as perpendicular to it - will be discussed in Section \ref{sec:outflow}.}

\item{{\bf Region E1:} Region E1 is the region at the edge of the weak radio emission about 4 kpc east of the nucleus. In region E1 the [\SII] emission line shows a broad component with FWHM = $708 \pm 80$ km~s$^{-1}$. In this case, the centre of this
broad component is blueshifted by only $114 \pm 48$ km~s$^{-1}$ w.r.t. the
narrow component.  Thus, also this component may represent a gas outflow,
although not as pronounced as in region E0.}

\item{{\bf Region E2:} A very remarkable feature can be seen in region E2 
at $\sim 7$ arcsec east of the nucleus. In Figure \ref{fig:overlay} this
feature clearly has a ``comma''-like shape and the ionized gas disk seems distorted. This is strengthened by the fact that in region E2 we cannot identify the narrow component in the [\SII] line. Instead, we fit a single broad component in region E2. The FWHM of this
feature is $664 \pm 27$ km~s$^{-1}$. When extrapolating the velocity that the
disk would have in region E2 (as traced by the narrow component in the regions
adjacent to E2; see Figure \ref{fig:artSII93b}-A), the centre of this broad
component is blueshifted by $202 \pm 22$ km~s$^{-1}$, again indicating the
presence of gas outflow also in this region. In region E2 there is no
detection of radio continuum down to a limit of 3 mJy~beam$^{-1}$ in radio
continuum maps presented by e.g. \citet{bri81}.}

\item{{\bf Regions W1 $\&$ W2:} Regions W1 and W2 are the regions west of the nucleus. In regions W1 and W2 the emission-line
    profile, although very faint, is clearly asymmetric. By including an
    additional, broader component we get an excellent fit to the emission-line
    profile in this part of the galaxy. The narrow component on the western
    side nicely traces the velocity curve of the gas disk, while the
    broader component is {\sl redshifted} with respect to this narrow component. However, the FWHM of the fit to this broader component in regions W1 and W2 is only $398
    \pm 68$ (W1) and $264 \pm 40$ (W2) km~s$^{-1}$. Furthermore, the shift
    between the narrow and broad component is only $115 \pm 75$ (W1) and $152
    \pm 33$ (W2) km~s$^{-1}$. This redshifted component is not as pronounced as the blueshifted component seen on the eastern side of the galaxy. It might, therefore, indicate
    that only a mild outflow is present on the western side of the galaxy
    (with gas moving away from us). Another possibility is that this component
    might be caused by line-of-sight effects when we look through the almost
    edge-on disk of the galaxy. The asymmetric emission-line profile would then represent the narrow component gas at various locations in the disk, intercepted by our line-of-sight. With the data available to us we cannot disentangle these two possibilities.}

\end{itemize}

{\sl The main result from this first part of the kinematic study of the
ionized gas is that there are a number of locations along the position angle
of the inner radio jet where gas outflows are observed.  In addition to this,
the most prominent outflow is observed at the location of the stronger radio
continuum (E0). This will be further discussed in Section \ref{sec:regionE0}.}

\subsection{Extended gas disk}
\label{sec:disk}

\begin{figure}
\includegraphics[width=8cm]{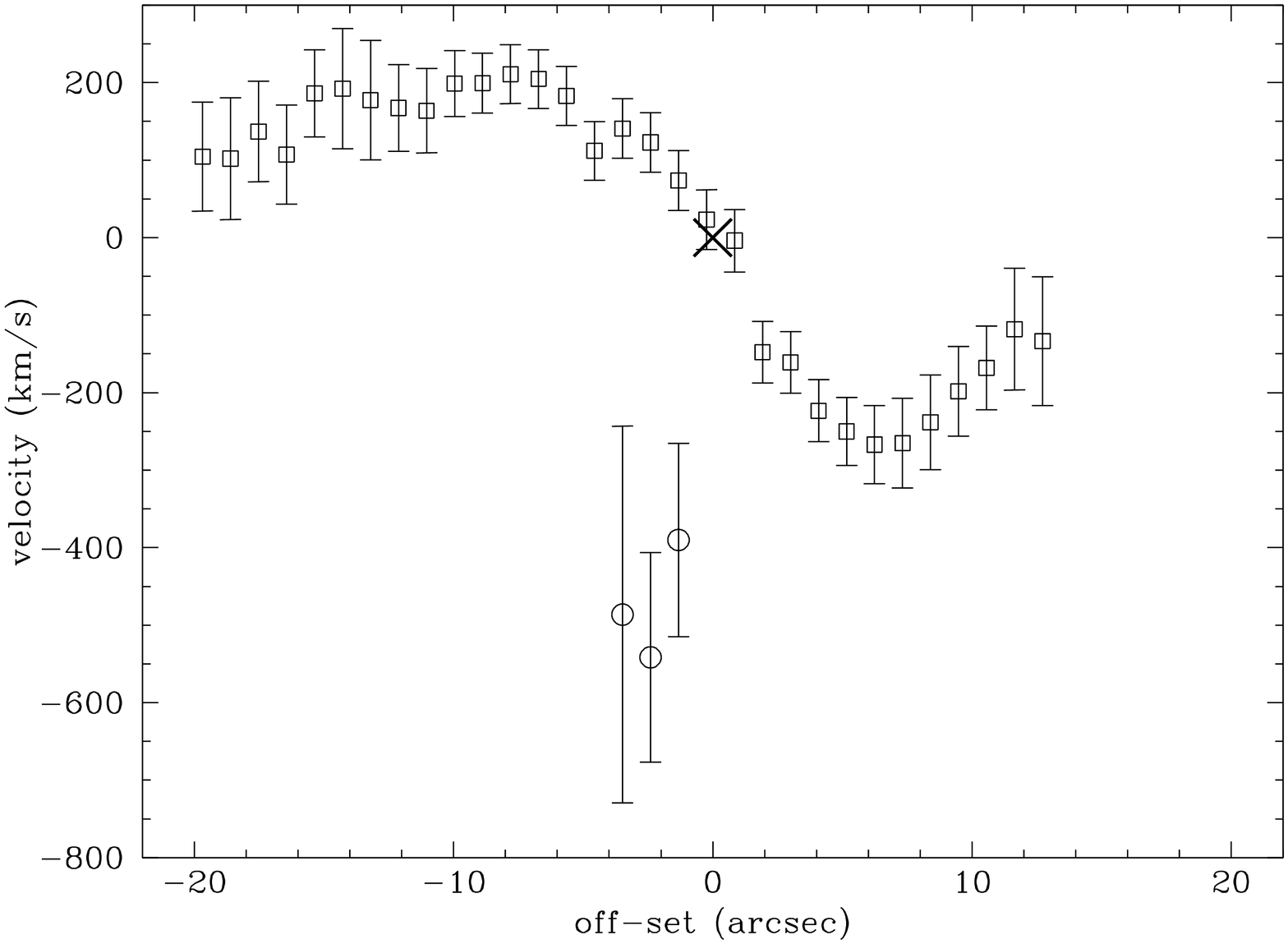}
\caption{Position-velocity diagram of the [\OII] line along the major axis of the galaxy (p.a. 60$^{\circ}$). The squares trace the narrow component; the circles the broad component. Point (0,0) represents the nucleus. The off-set on the horizontal axis is the off-set (in arcsec) from the nucleus (error $\pm$ 0.6 arcsec). On the vertical axis the velocity is plotted w.r.t. the systemic velocity that we derive at the nucleus (v$_{sys}$=13,450 $\pm$ 35 km s$^{-1}$; see Section \ref{sec:radioaxis}). The vertical error bars represent uncertainties in the Gaussian fits and inaccuracies in the wavelength calibration.}
\label{fig:curveOIIart1b}
\end{figure}

The spectrum taken along p.a. 60$^{\circ}$ follows the major axis of the
galaxy. Therefore we also analysed the kinematics of the emission-line gas
in this direction. Figure \ref{fig:curveOIIart1b} show a velocity-curve of the ionized gas, constructed by fitting Gaussian profiles to the [\OII] line, using apertures of 0.6 arcsec at various places along the 60$^{\circ}$ slit. Being one of the stronger lines, [\OII] can be traced out to a larger distance from the centre than other lines. From this diagram it is clear that the [\OII] emission-line gas follows
the kinematics of a regular rotating disk \citep[also seen by][]{bre84}. The
plot also clearly shows how the blueshifted component observed in region E0
(the only one included in this slit position) is well separated from the
gravitational motion of the regular rotating disk gas. The total diameter of the gas disk is 30 kpc, which is larger than the 19 kpc disk already found by \citet{bre84}. A small difference in slit alignment might explain the slight deviation of our velocity curve from the one of \citet{bre84}. 

We centre the velocity curve on the systemic velocity that we derived in Section \ref{sec:radioaxis} ($v_{\rm sys} = 13,450 \pm 35$ km~s$^{-1}$). Our value for $v_{\rm sys}$ is in agreement with $v_{\rm sys} = 13,478$ km~s$^{-1}$, found with \HI\ absorption at the core by
\citet{bes04}. The redshift velocity of the disk gas in region E0 is roughly
13,500 km~s$^{-1}$, consistent with the velocity of the deep, narrow \HI\
absorption that is located against the enhanced radio emission in region E0
\citep{bes02,bes04} and with CO measurements \citep{eva99}.
From our rotation curve along p.a. 60$^{\circ}$ we derive a velocity gradient
of $\sim 57$ km~s$^{-1}$~arcsec$^{-1}$ in the inner region of the galaxy. This
is consistent with the gradient seen by \citet{bre84} and with the gradient of $\sim 46$ km~s$^{-1}$~arcsec$^{-1}$ derived from \HI\ absorption studies by \citet{bes02}.\\

\subsection{Ionization and density of the gas}
\label{sec:ionization}

Table \ref{tab:ratioshotspot} gives the values of the fluxes of the
narrow component of various emission lines measured in the regions E0 and C
in the spectrum along p.a. 93$^{\circ}$ (see Figure \ref{fig:specartfin2}). The fluxes are the total emission-line fluxes, estimated from Gaussian fits, across an aperture of 0.8 arcsec (corresponding to a spatial binning of 4 consecutive pixels in the 2D-spectra) centred on regions E0 and C.

The line-ratios in Table 3 are not corrected for extinction. The ratio
H$\alpha$/H$\beta$ = 9.5 in region E0 is about a factor 3 too high when comparing it with the hydrogen recombination analysis done by \citet{ost89}. Therefore it is clear that \Hb\ is strongly
affected by reddening \citep[caused by various dust lanes that cross the inner
region of 3C~293, e.g.][]{mar99} and/or by underlying \Hb\ absorption associated
with the young stellar population that \citet{tad05} found in this
galaxy. Furthermore, the \Ha\ emission line could be affected by underlying
atmospheric absorption.  All this makes the analysis of the line ratios more
complicated. Nevertheless, when comparing with diagnostic diagrams \citep[e.g.][]{bal81,vei87,dop95}, we can use some of the lines, which lie
close together so that the line ratios are not affected by extinction 
too much (e.g. [\NII]/H$\alpha$, [\SII]/H$\alpha$ and [\OI]/H$\alpha$ ratios, 
together with [\OIII]/H$\beta$), as good diagnostics for the classification. 
Using these diagnostics we can conclude that region E0 in 3C~293 has a LINER-type 
(Low Ionization Nuclear Emission-line Region) spectrum \citep{hec80}. Thus, the 
narrow component gas in region E0 is likely either shock ionized or 
photo-ionized by the AGN at low ionization parameter. The line ratios are inconsistent 
with photo-ionization by O and B stars (\HII\ region) or by a strong power-law AGN component at
high ionization parameter.

The [\OII] and [\SII] fluxes of the broad component gas associated with the
outflow in region E0 are also listed in Table \ref{tab:ratioshotspot}. We were
not able to make an accurate fit of the broad component in region E0 to any
other line than the strong [\OII] and [\SII] lines. Other lines also show
evidence for a broad component, but due to the underlying continuum and the
limited signal-to-noise (in combination with the complexity of some of the
emission lines, like e.g. [\NII]+H$\alpha$) we were not able to get a
reliable fit to these broad components.

The emission-line gas in region C (fitted with a single Gaussian component) appears to have ionization characteristics similar to the narrow line
gas in region E0, although the measured H$\alpha$/H$\beta$ = 15.7 indicates
that region C is likely even more affected by dust-extinction than region
E0. In region C the gas is also likely ionized either by shocks or by the AGN at low ionization parameter.

\begin{table}
\caption{Fluxes of the narrow component emission-line gas - relative to \Hb\ - across an 0.8 arcsec aperture in regions E0 and C. The H$\beta$ flux is given in erg s$^{-1}$ cm$^{-2}$. The errors are also scaled to the value of the \Hb\ flux and reflect both the uncertainty of the Gaussian fits as well as the inaccuracy in determining the continuum level. Also included are the fluxes of the broad component of the [\OII] and [\SII] gas (scaled to the value of the narrow component) in region E0.}
\label{tab:ratioshotspot}
\begin{tabular}{l|c|c}
 & {\bf Region E0} & {\bf Region C} \\
Element & Flux &  Flux \\
(narrow comp.) & relative to H$\beta$ & relative to H$\beta$ \\
\hline   
\hline  
H$\beta$ flux & 2.32$\pm$0.24$\times$10$^{-16}$ & 1.15$\pm$0.10$\times$10$^{-16}$ \\ 
\hline
$[\rm \OII]$$\lambda$3727 & 4.2 $\pm$0.2 & 4.9 $\pm$ 0.1 \\
$[\rm \OIII]$$\lambda$4363 & $<$0.3 & - \\ 
$[\rm \OIII]$$\lambda$5007 & 1.2 $\pm$ 0.2 & 2.6 $\pm$ 0.1 \\ 
$[\rm \OI]$$\lambda$6300  & 1.2 $\pm$0.4 & 5.2 $\pm$ 0.1  \\  
H$_{\alpha}$  & 9.5 $\pm$0.4 & 15.7 $\pm$ 0.7 \\
$[\rm \NII]$$\lambda$6583 & 7.7 $\pm$2.9 & 20.4 $\pm$ 3.8 \\
$[\rm \SII]$$\lambda$6716+$\lambda$6731 & 7.4 $\pm$0.3 & 21.9 $\pm$ 0.3 \\
\hline
$[\rm \OII]$$_{\rm broad}$/$_{\rm narrow}$ & 0.32 $\pm$0.05 & - \\
$[\rm \SII]$$_{\rm broad}$/$_{\rm narrow}$ & 0.30 $\pm$0.05 & - \\
\end{tabular}
\end{table} 

\begin{figure}
\includegraphics[width=8.5cm]{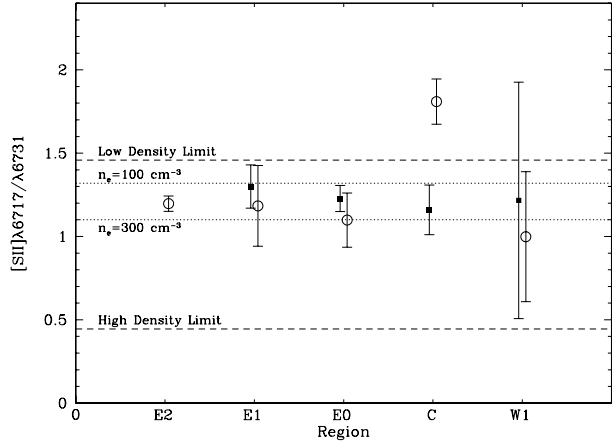}
\caption{[\SII]$\lambda$6716/$\lambda$6731 of both the narrow emission-line gas component (closed squares) as well as the broad emission-line gas component (open circles) in the various regions along p.a. 93$^{\circ}$. Shown are the dotted lines that correspond to densities of 100 and 300 cm$^{-3}$ \citep[assuming $T = 10,000$ K, see][]{ost89} as well as the maximum and minimum value for the [\SII] doublet line ratio that is physically possible (the upper dashed line corresponds to $n_{\rm e} \leq 10^{1}$ cm$^{-3}$, the lower dashed line to $n_{\rm e} \geq 10^{5}$ cm$^{-3}$. The error bars reflect the uncertainty in fitting the Gaussian components.}
\label{fig:densityplot}
\end{figure}

Electron densities of the gas can be derived from the [\SII]$\lambda$6716/$\lambda$6731 ratio \citep{ost89}. Figure \ref{fig:densityplot} shows [\SII]$\lambda$6716/$\lambda$6731 for both the narrow and the broad component in the various regions along p.a. 93$^{\circ}$ (the fit is too uncertain across region W2 and therefore we leave it out of the plot). The broad component represents different features in the various regions (Section \ref{sec:radioaxis}), which might be physically unrelated. Nevertheless, given the relatively large uncertainties, we do not see any difference in the [\SII] ratio in the various regions.  The weighted mean
[\SII]$\lambda$6716/$\lambda$6731 ratio in Figure \ref{fig:densityplot} is
$1.23 \pm 0.08$ for the narrow component and $1.19 \pm 0.07$ for the broad
component (excluding the broad component in region C, that does not represent
an outflow; see Section \ref{sec:radioaxis}). This corresponds to densities of
about $2 \pm 1 \times 10^{2}$ cm$^{-3}$ for the narrow component and $3 \pm 1
\times 10^{2}$ cm$^{-3}$ for the broad component gas, assuming $T = 10,000$ K
(since [\OIII]$\lambda$4363 is not detected we are not able to derive the
electron temperature of the emission-line gas). Thus, we do not see any
significant difference between the density derived for the narrow and the
broad component gas.

\section{Fast gas outflow in region E0}
\label{sec:regionE0}

The fastest outflow of gas is seen in region E0. In this region the
emission lines are particularly strong and they allow some further
analysis of the characteristics of this outflow. In particular we
investigate in this Section the spatial extent of the region showing
the fast outflow and we estimate the mass of the outflowing gas. This information will be used later on for building a possible scenario about the origin of
this outflow.

\subsection{Spatial extent of the gas outflow in region E0}
\label{sec:outflow}

\begin{figure*}
  \includegraphics[width=17.5cm]{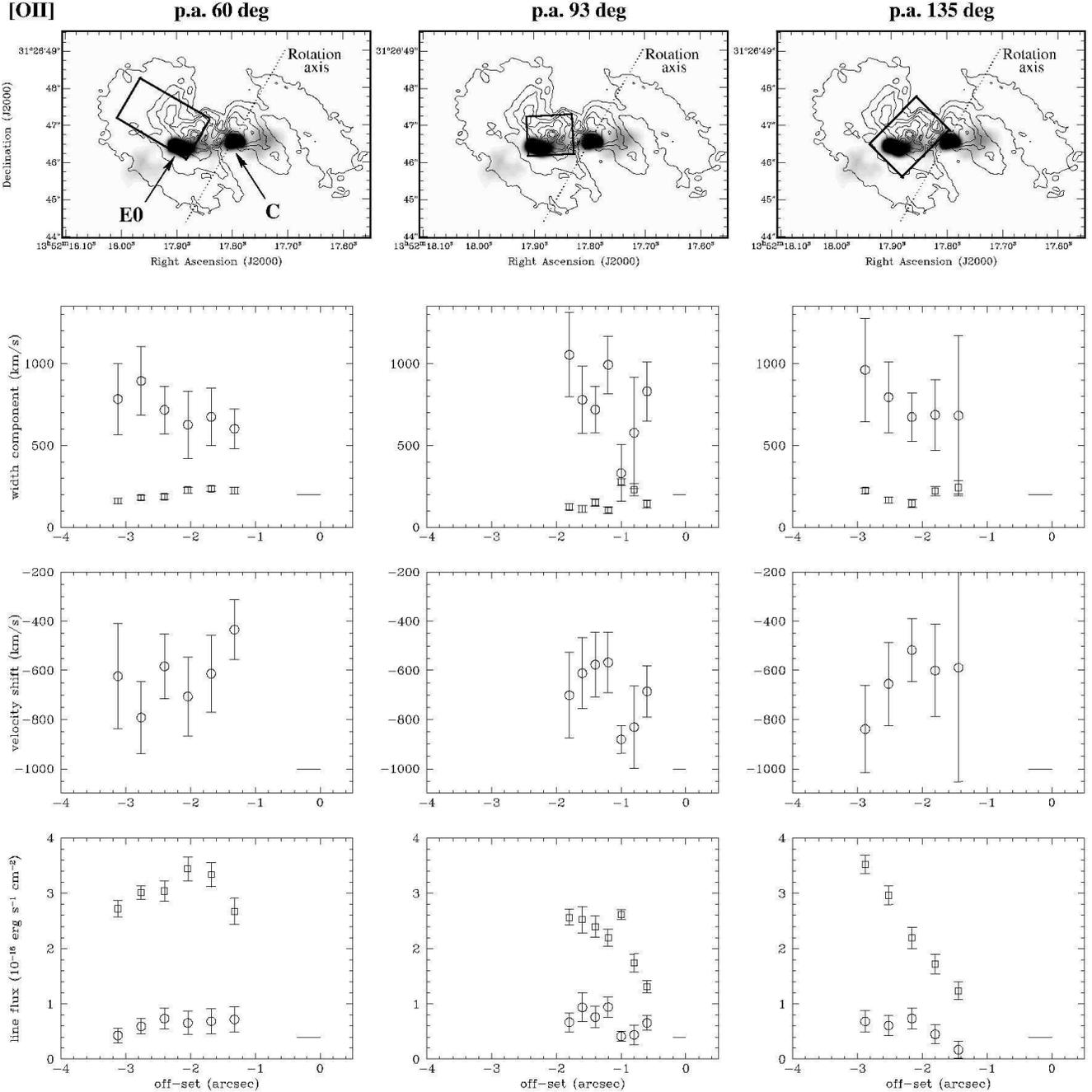}
\caption{In the top plots contours of the HST F702W image of the inner region of 3C~293 are overplotted onto a grey-scale plot of the MERLIN radio continuum map. The dotted line presents the rotation axis of 3C~293 through the nucleus. The boxes indicate the region in which the broad, blueshifted component of the [\OII] emission line is detected (see text for the accuracy of the position of the boxes). In the plots below that, various observed parameters are plotted. The first plot shows the width of the broad (circles) and the narrow (squares) component (corrected for instrumental broadening). The plots below that show the velocity shift between the broad and narrow component. The bottom plots show the line flux of both the broad (circles) and narrow (squares) component. The off-set on the horizontal axis is w.r.t. the nucleus, which we can determine with an uncertainty of 0.6 arcsec for the [\OII] line. The horizontal bars in the bottom-right corner of the plots indicate the apertures over which we summed rows of pixels to extract the 1D-spectra (note that for the line flux along p.a. 60$^{\circ}$ and 135$^{\circ}$ we used a 0.4'' aperture and a 1.3'' slit width, while for the p.a. 93$^{\circ}$ the aperture was only 0.2'' and slit width 1.03'').}
\label{fig:extendOIInew3}
\end{figure*}

In Figure \ref{fig:extendOIInew3} we analyse the extent of the region in which we 
detect the fast outflow. We do this by tracing the broad, blueshifted component 
of the [\OII] emission line along the different position angles (since the [\OII] 
line is stronger than the [\SII] line it provides a better indication for the presence 
of the broad component). The top plots show in contours the HST F702W
image of the central few kpc of 3C~293. The radio continuum of the MERLIN map
\citep[from][]{bes02} is plotted in grey-scale and clearly shows the knots of
enhanced radio emission in the regions C and E0. The length of the boxes indicates
the region in which the broad, blueshifted component is detected; the width of
the boxes corresponds to the width of the slit. Note that the slit was aligned
by eye to cover the brightest region and Figure \ref{fig:extendOIInew3} merely
shows the most likely positioning of the slit. The uncertainty in the position
of the nucleus in the spectra around the [\OII] line (the ``0''-point 
for the spatial off-set) is 0.6 arcsec. Also shown in Figure \ref{fig:extendOIInew3} 
are the width of the broad and narrow component, the velocity shift between the two 
components and the line flux of both components across the region in which the broad component is detected.

From Figure \ref{fig:extendOIInew3} and the seeing listed in Table
\ref{tab:obsparam} it is clear that the region in which the broad, blueshifted component 
is observed, i.e. the fast-outflow region, is spatially
resolved. Not only is it resolved along the direction of the inner 
radio axis (where, moreover, we know already other regions of gas outflow do exist; see 
Section \ref{sec:radioaxis}), but also perpendicular to it. The gas outflow 
is observed in a region of $2 \times 1.5$ kpc$^{2}$.

\subsection{Mass of the gas outflow in region E0}
\label{sec:mass}

The mass of the gas in a line emitting region is related to the \Hb\ luminosity in that
region by:
\begin{equation}
M_{\rm gas}=m_{\rm p}{{L(\rm H\beta)}\over{n_{\rm e}\alpha^{\rm eff}_{\rm H\beta}h\nu_{\rm H\beta}}}
\label{eq:masslhb}
\end{equation}
\citep[see][]{ost89}, where $n_{\rm e}$ is the electron density (cm$^{-3}$),
$m_{\rm p}$ the mass of a proton (kg), $L(\rm H\beta)$ the \Hb\ luminosity
(erg s$^{-1}$), $\alpha^{\rm eff}_{\rm H\beta}$ the effective recombination
coefficient for \Hb\ (cm$^{3}$ s$^{-1}$) and $h\nu_{\rm H\beta}$ the energy of
a \Hb\ photon (erg). Assuming that the fast outflow is spread over an area of $2
\times 1.5$ kpc$^{2}$ (Section \ref{sec:outflow}) and accordingly over 1.5 kpc
in our line-of-sight direction, we first estimate the luminosity of the narrow
component gas of the disk in this outflow region to be $L(\rm H\beta)_{\rm narrow} = 1.3 \times 10^{40}$ erg$~$s$^{-1}$. Here we assume that the extinction corrected flux (about $3 \times$ the flux given in Table \ref{tab:ratioshotspot}, which was measured with a slit aperture of $0.8 \times 1.03$ arcsec$^{2}$; see Section \ref{sec:ionization}), uniformly covers the region of outflow. This is a reasonable assumption, as can be seen from Figure \ref{fig:extendOIInew3} (bottom). To estimate the mass of the outflowing gas we need $L$(\Hb)
of the broad component gas. Unfortunately, the quality of the data did not
allow us to unambiguously fit a broad component to the \Hb\ line (see Section
\ref{sec:ionization}). Nevertheless, when assuming that the line ratio
\Hb$_{broad}$/\Hb$_{narrow}$ $\sim 0.3$ (similar to what we measured for [\OII]
and [\SII]; Table \ref{tab:ratioshotspot}), we derive for the \Hb\
luminosity of the broad component $L(\rm H\beta)_{broad} \approx 4.3 \times
10^{39}$ erg~s$^{-1}$.  Using the density derived in Section \ref{sec:ionization}, we
estimate that the {\sl total mass of the outflowing ionized gas in region E0 is $M \approx
1 \times 10^{5}$ M$_{\odot}$}.  The total ionized gas mass for the
narrow component disk gas in the outflow region is $M
\approx 6 \times 10^{5}$ $M_{\odot}$ 

With the oversimplified assumption that the gas is locked up in clouds with a derived density of $3 \times 10^{2}$ cm$^{-3}$ one can make an estimate for the filling factor ($F_f$) of these clouds over the region where we detect the fast outflow:
\begin{equation}
F_f = {{M}\over{n m_{\rm p} V}}
\label{eq:massdirect}
\end{equation}
where $M$ is the total mass of the outflowing gas, n the density of the gas, m$_{\rm p}$ the mass of a proton and $V$ the volume of the region where the fast outflow is detected. Using the derived properties of the gas involved in the fast outflow, Equation \ref{eq:massdirect} gives a filling factor of the outflowing gas of $F_{f}$ $\approx$ 3 $\times$ 10$^{-6}$. Of course this is a rough estimate, since the distribution of the gas is unknown, but it is consistent with values derived in other powerful radio galaxies \citep{bre85,bre86,hec89,koe98}. It shows that we are dealing with a clumpy medium in the central region of 3C~293.

\section{Comparison with the \HI\ outflow}
\label{sec:hi}

\begin{figure*}
\includegraphics[width=17cm]{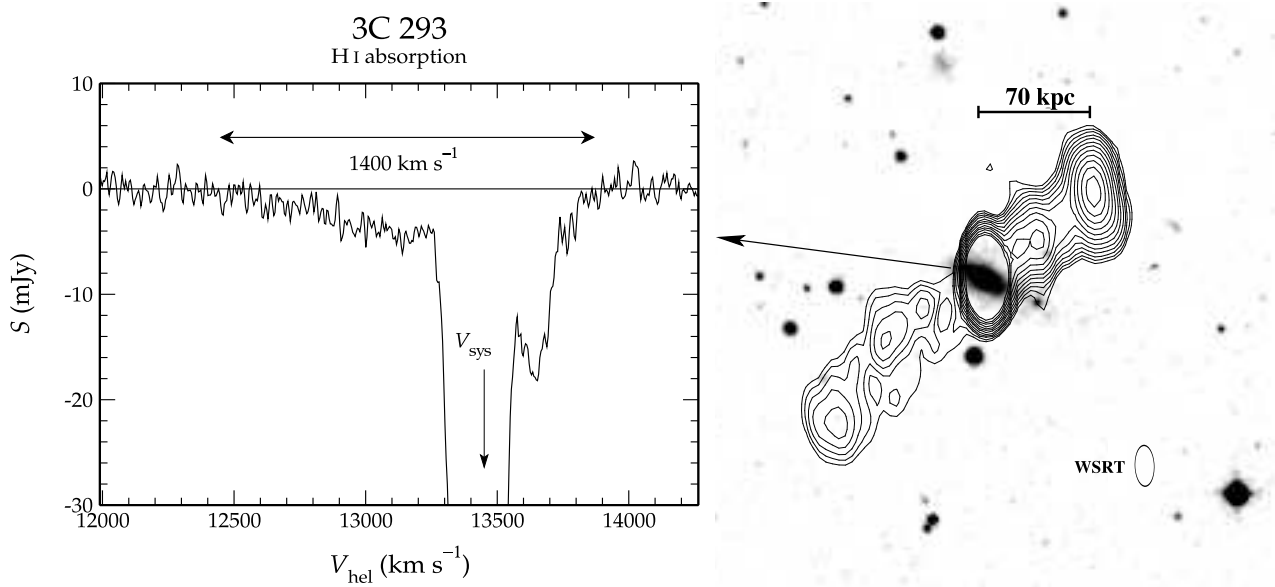}
\caption{Right: Continuum image of 3C~293 made with our 10 MHz data. Left: Zoom-in on the \HI\ absorption spectrum detected against the unresolved central radio continuum using the 20 Mhz band (Figure from Paper 1). Indicated is the full width at zero-intensity of the \HI\ outflow.}
\label{fig:hiab}
\end{figure*}

In Paper 1 we presented evidence for a fast outflow of neutral hydrogen gas in 3C~293. 
Figure \ref{fig:hiab} (from Paper 1) shows the broad but shallow \HI\ absorption, mainly 
blueshifted w.r.t. the systemic velocity, that represents this fast outflow of \HI. This 
absorption is detected against the central radio continuum using the 20 MHz band at the WSRT, but due to the low spatial 
resolution of the WSRT we are not able to locate the region in the inner radio structure 
from which the \HI\ outflow originates.

\begin{figure}
\includegraphics[width=8cm]{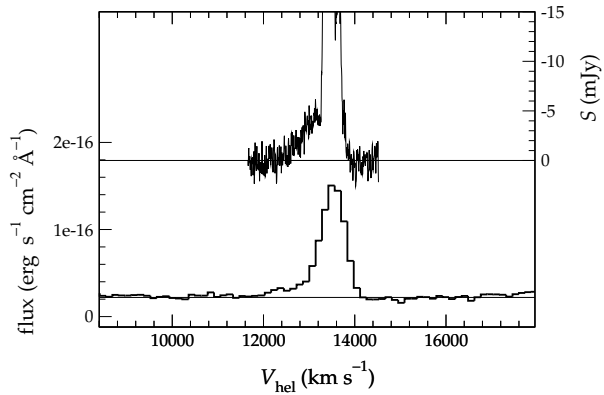}
\caption{[\OII] emission line in region E0 and \HI\ absorption profile (inverted) plotted in one image.}
\label{fig:hioiiplot}
\end{figure}

In \citet{emo04} we made a Gaussian fit to this broad \HI\ absorption and
derived a FWHM of $852 \pm 41$ km~s$^{-1}$. The centre of this Gaussian fit is
$345 \pm 24$ km~s$^{-1}$ blueshifted from the deepest, narrow \HI\ absorption
feature, which traces quiescent gas likely located in an extended \HI\ disk \citep{has85,bes02,bes04}. The characteristics of the broad, blueshifted \HI\ component look similar to
those of the broad component of the ionized gas observed in region E0. A
comparison between the two is shown in Figure \ref{fig:hioiiplot} (from
Paper 1). Region E0 therefore appears to be the most likely location for the fast
\HI\ outflow.  The fact that the radio continuum is strongest in region E0 further strengthens this conclusion. The observed outflow of \HI\ cannot
be associated with the outflow of ionized gas that we see in region E2, since
no strong radio continuum is detected in region E2. Moreover, at the
resolution of the WSRT observations, the  outflow would appear not coincident
with the nucleus but displaced with respect to it.

{\sl The similarity between the characteristics of the \HI\ outflow and the ionized gas outflow in region E0 suggests that they likely originate from the same driving mechanism.}

If the broad component of the \HI\ absorption is indeed located against
the region E0, we can derive a new value for the \HI\ column density. The peak
intensity of the bright radio emission in region E0 in the MERLIN map of
Figure \ref{fig:slitposition} is 1.29 Jy.  Therefore, the optical depth that
the broad \HI\ absorption feature has if it is located in region E0 is $\tau =
0.38\%$.  This corresponds to column densities of the outflowing \HI\ gas of $\sim 6 \times 10^{20}$ cm$^{-2}$ assuming $T_{\rm spin} = 100$ K.  This value
for the column density likely represents a lower limit, as in the extreme conditions under which these fast outflows occur the $T_{\rm spin}$ is likely (much) higher \citep[e.g.][]{bah69,mal96}. 

Assuming that the
outflowing \HI\ gas extends over the same region as the fast outflow of ionized
gas, the \HI\ column in front of the radio plasma will extend over $\sim 0.75$
kpc. The mean density of the neutral gas involved in the fast outflow will
therefore be $\sim 0.3$ cm$^{-3}$, which is comparable to the densities of $\sim
0.1$ cm$^{-3}$ that \citet{has85} found for \HI\ clouds that are falling onto
the AGN. However, given the very low filling factor that we derive for the
ionized gas (Section \ref{sec:mass}), the neutral gas will most likely also be
locked up in a clumpy medium with locally much higher densities. If the
outflowing \HI\ gas extends over the same region as the fast outflow of
ionized gas in region E0 we derive a total \HI\ mass of $M_{\rm \HI} \approx
10^{7}$ $M_{\odot}$  involved in the fast outflow. Although
uncertainties are large (mainly due to the uncertainty in the exact location
and extent of the \HI\ outflow and due to the unknown spin temperature of the
gas), the bulk of the outflowing gas nevertheless appears to be in a cold,
neutral state and only a minor fraction is ionized.

\section{DISCUSSION}
\label{sec:discussion}

The long slit spectra presented in this paper show complex kinematics
of the ionized gas in the radio galaxy 3C~293, in particular along the position angle of the inner radio jets. At a number of locations we find outflows of ionized gas. The broader and most blueshifted gas-component, that we identify with a fast ($\sim 1000$ km~s$^{-1}$) outflow of ionized gas, is detected in the region E0, close to the brighter radio emission found along the jet.  This is intriguing and suggests that the presence of the radio
jet may play a role in producing the outflow.  The total mass of warm gas
involved in the fast outflow is $\sim 1 \times 10^{5}$ $M_{\odot}$.  There are
more regions in which complex kinematics of the ionized gas are seen, although
less extreme than what we detect in region E0. 

In addition to this, one of the most interesting results obtained is the
similarity in the kinematic properties between the ionized gas outflow in the region E0 and the outflow of neutral hydrogen presented in Paper 1. {\sl This suggest that the two outflows are coming from the same region (that would therefore be
about 1 kpc from the nucleus) and that they are likely driven by the same
mechanism.} The total \HI\ mass of $10^{7}$ $M_{\odot}$ involved in the outflow shows that most of the outflowing gas is in its neutral state and that only a small fraction is ionized.

In the following, we will discuss some of the possible mechanisms that can cause such (off-nuclear) outflows, focusing in particular on the
case of the region E0, for which detailed information could be derived from the
optical data. Although the scenario of a jet-ISM interaction appears to be the
most likely mechanism for the fast outflow in region E0, other mechanisms are
also possible. We also briefly discuss what could be the cause for the milder outflows observed in other regions across 3C~293.

\subsection{Driving mechanism for the fast gas outflow in region E0}
\label{sec:drivingmechanism}

Here we consider three possible mechanisms to explain the detected gas
outflows, in particular the fast outflow detected in region E0.
 
\begin{itemize}

\item{{\bf Starburst driven wind:} 

A possible driving mechanism for the
outflow is acceleration of the gas by winds from young stars.  By fitting the
Spectral Energy Distribution (SED) at various places in 3C~293, \citet{tad05}
detected a young/intermediate age stellar population of 0.1 - 2.5 Gyr
throughout 3C~293. Although starburst winds can generally produce outflows up to 1000
km~s$^{-1}$, can easily carry enough energy for an outflow such as observed in
3C~293, and can predominantly shock-ionize the gas \citep[e.g.][]{hec90}, it is not clear 
whether a starburst wind can survive over a period of 0.1 - 2.5 Gyr. Even if this is
 possible, a wind from this young/intermediate age stellar population would have to be ``fossil'' in order to still be visible at present. An outflow driven by such a fossil wind 
 would no longer be localized to the region E0 where we see it, but would be many tens of kpc in extent.

It should be noted, however, that bright complexes of UV emission have been 
found throughout the disk of 3C~293 \citep{all02}. Some of these complexes could contain 
young O and B stars, in which case winds from these young stars may contribute in some regions to the 
complex kinematics of the gas that we observe. The relatively poor accuracy of the astrometry between the 
UV image and the radio data does not allow us to reliably estimate how these complexes are 
located with respect to the radio features. It is worth noting, however, that one of the brighter 
UV complexes (one with an intriguing {\bf V}-shaped morphology) could be located close to the region E0.}\\

\item{{\bf AGN radiation:} 

\cite{kro86} and \cite{bal93} suggest that the continuum from the central 
engine can cause X-ray heated winds, which contain gas that evaporated off a 
surrounding torus. Their models show that the flow speeds of these winds can reach several 
hundred km s$^{-1}$ in the pc-scale region around the AGN. Following \citet{dop02}, also further from the nucleus the nuclear radiation can be responsible, if coupled to dust, for radiation pressure dominated gas outflows. This idea has been applied to the case of NGC~1068, where it can explain an outflow of $\sim
700$ km s$^{-1}$ at about 100 pc from the nucleus. A similar mechanism has been suggested in the case of Cygnus A \citep{bem03}. In this case, the radiation from the powerful AGN in this radio
galaxy can explain the outflow of $\sim 170$ km s$^{-1}$ observed in the ionization cone at about 1 kpc from the nucleus.

However, the AGN in the core of 3C~293 appears to be relatively weak compared
with the objects considered above. Although obscuration certainly plays a role,
we nevertheless find that the warm gas has a low ionization state in the centre of
3C~293. In particular, the [\OIII]$\lambda$5007 line is  weak.
Alternatively, we can use the far-IR luminosity and assume that it is all due
to re-radiated quasar-light instead of starlight. The far-IR luminosity of
3C~293, $L_{\rm 60 \mu m}/L_{\odot} = 10.08$ \citep{gol88}, is at the lower end of the
luminosity for quasars \citep{neu86}. This is more than an order of magnitude
lower than for the nucleus of Cygnus A and about a factor 6 lower than for the nucleus of NGC~1068. 
It seems, therefore, unlikely that the radiation from the nucleus of 3C~293 is strong enough to accelerate the gas to the large velocities observed in region E0. Also in the case of Cygnus~A, 
the much faster ($\sim$1800 km s$^{-1}$) outflow that has been detected along the radio axis in the 
[\OIII] emission lines \citep{tad91} is not likely driven by AGN radiation pressure, but more likely explained as jet-induced.}\\

\item{{\bf Jet induced outflow:} 

The fastest outflow of gas is seen $\sim$1 kpc east of the nucleus in a region (E0) 
where there also is enhanced radio continuum due to a propagating radio jet. At 
VLBI-scale \citep[e.g.][]{bes04} this radio jet shows a distorted morphology, including 
several bright knots, in particular near region E0. \citet{bes04} argue that 
the radio jet in region E0 is approaching and that the continuum emission in this region 
might not be related to the fast jet itself, but rather to a low-velocity shear layer surrounding the jet, 
possibly created by an interaction between the jet and the ISM. In the central few kpc of 
3C~293 there is a large reservoir of dense, cold gas, including HI 
clouds \citep[with individual cloud masses $\la$ few times $10^{6}$ $M_{\odot}$;][]
{has85,bes02} as well as $1.5 \times 10^{10}$ $M_{\odot}$ of H$_{2}$ \citep{eva99}. Overall these results suggest that the observed fast outflow of gas is caused by an 
interaction between the propagating radio jet and this dense ISM. Moreover, 
the outflow in 3C~293 resembles cases of jet-related outflows in other nearby 
powerful radio galaxies like Cygnus A \citep{tad91}, PKS 2250-41 \citep{vil99}, 
PKS 1549-79 \citep{tad01}, PKS 1345+12 \citep{hol03} and 3C~305 \citep{mor05}.

The total mass of warm gas involved in the outflow is $\sim 1
\times 10^{5}$ $M_{\odot}$, while the mass of the outflowing HI gas is $\sim 10^{7}$ $M_{\odot}$. We can estimate a mass outflow-rate of the gas, assuming that
the outflow is free-floating and spherical in shape around the radio knot (we assume that the outflow
covers at least 2$\pi$ steradians as seen from the jet at the origin of the outflow):
\begin{equation}
\dot{M} \approx 2 \pi R^2 \cdot \rho \cdot F_f \cdot v_{\rm out} \cdot m_{\rm H},  
\label{eq:massoutflow}
\end{equation}
with $R$ the radius of the outflowing spheroid of gas, $\rho$ the volume density and
$F_{f}$ the filling factor of this gas, $v_{\rm out}$ the velocity
of the outflowing gas and $m_{\rm H}$ the mass of a H-atom. For the gas in region E0 that 
is flowing out at a velocity of 1000 km~s$^{-1}$ in a
spherical shell with radius $R = 0.75$ kpc we determine $\dot{M} \sim 0.1$
$M_{\odot}$~yr$^{-1}$ for the ionized gas and $\dot{M} \sim 23$ $M_{\odot}$~yr$^{-1}$ for the \HI\ gas. 
Even if the inner radius of the spheroid of outflowing \HI\ is not larger than the actual size of the radio knot \citep[r$_{\rm knot}$=32 pc;][]{aku96}, the \HI\ mass outflow-rate will still be of the order 1 $M_{\odot}$~yr$^{-1}$. These outflow rates are conservative estimates, since the spin temperature of the outflowing gas is most likely higher than the assumed 100 K. The calculated outflow-rates are nevertheless in agreement with the total mass of the outflow that we see at present. In case of a steady-state outflow of gas, the outflow must have been driven for about the past 10$^{6}$ yr in order to get a total outflow of $M_{\rm \HI} \sim 10^{7}$ $M_{\odot}$ and $M_{\rm ion} \sim 10^{5}$ $M_{\odot}$. During this time the outflowing gas has reached a distance of $\sim$1 kpc, which is also in agreement with the size of the region in which we detect the outflow. \citet{aku96} estimate the age of the outer radio structure to be $\sim 2.69 \times 10^{6}$ yr, assuming that the jet is moving freely at an advance speed of 0.1{\sl c}. Their estimate of the age of the inner radio jets (when due to re-started activity) is $\sim 2.95 \times 10^{4}$ yr, although we note that there must be a large uncertainly in this estimate, given the fact that the inner radio jet is most likely relativistic \citep{bes04} and interacting with a dense ISM. Nevertheless, our best estimate of 10$^{6}$ yr for the age of the outflow implies that the radio plasma has been in interaction with the gas for a significant fraction of the total lifetime of the radio source. A total mass outflow-rate of 23 $M_{\odot}$~yr$^{-1}$ is at the low end of galaxy-scale stellar winds observed by \citet{rup02} in a sample of Ultra Luminous Infra-red Galaxies (ULIGs), indicating that the jet-ISM interaction could be an important factor in the evolution of this galaxy.

\subsubsection{Details of the jet-ISM interaction}

Let us now focus on {\sl how} the gas is driven out by the radio jet, 
taking into account the energy-budget of the radio plasma. First we consider whether the 
radio power measured for 3C~293 is indeed sufficient to produce the 
observed outflow. The total power of 3C~293 is $P_{\rm 1.4 GHz} \sim 2 \times 10^{25}$ W~Hz$^{-1}$ (Section \ref{sec:observations}). Following \citet{bic02} (and references therein), the rate at which radiating electrons are supplied to the radio lobes is proportional to the energy flux ($F_{\rm E}$) of the radio plasma, so that:
\begin{equation}
\kappa_{\nu} = {{P_{\nu}}\over{F_{E}}}
\end{equation}
where $\kappa_{\rm 1.4 GHz}$ is normally assumed or derived to be in the order of 10$^{-12}$ - 10$^{-11}$ for radio galaxies on the tens of kpc scale. Assuming  $\kappa_{\rm 1.4 GHz} = 10^{-12}$, and assuming that half of the energy flux is deposited in the eastern and the other half in the western jet/lobe structure, we expect that $F_{\rm E}$ is in the order of 10$^{44}$ erg s$^{-1}$ for the eastern jet/lobe structure in 3C~293. From arguments mentioned in the previous Section, it is uncertain which part of the eastern radio jet/lobe structure is (and has been in the past) responsible for the fast outflow, and therefore also what the age of the outflow is. However, following the rough estimate from the previous Section that the outflow has been driven for about the past 10$^{6}$ yr, the total energy supplied by the radio jet in region E0 over this period is in the order of a few $\times$ 10$^{57}$ erg. The total kinetic energy of the outflowing gas in region E0 is $E_{\rm kin} = {{1}\over{2}}Mv^{2} \approx 1 \times 10^{56}$ erg. We therefore argue that the radio plasma in 3C~293 carries enough energy to create the fast outflow in region E0.  

The gas is accelerated by shocks that are created by the interaction between the jet and the surrounding medium. The gas outflow due to these shocks can be either momentum- or energy-driven. Let us first consider the momentum-driven case, in which the acceleration of the gas is caused by the impact of the working surface of the jet directly on a cloud. This happens when the jet is much lighter than the cloud. In this case only a small fraction of the jet energy flux is transmitted to the cloud, the rest is advected with the jet. The momentum-flux ($F_{p}$) of a jet is only a small fraction of its energy-flux \citep{bic02}: 
\begin{equation}
\begin{array}{ll}
F_{p} = \displaystyle{{1}\over{c}} \times F_{E} & (\rm relativistic) \\
 & \\
F_{p} = \displaystyle{{2}\over{v_{\rm jet}}} \times F_{E} & (\rm non-relativistic), \\
\end{array}
\label{eq:momentum1}
\end{equation}
For 3C~293 we saw that $F_{E} \approx 10^{44}$ erg s$^{-1}$ for the eastern jet/lobe structure. \citet{bes04} argue that the inner radio jet is most likely relativistic, therefore $F_{p} \sim 3 \times 10^{33}$ g cm s$^{-2}$. We can also derive the total momentum-flux of the outflowing gas directly from our observations:
\begin{equation}
\begin{array}{l}
F_{p'} = \dot{M} \times v \\ 
\ \ \ \ \ \sim 6 \times 10^{34} \left({\dot{M}}\over{10\ M_{\odot}\ {\rm yr^{-1}}}\right)\left({v}\over{1000\ {\rm km\ s^{-1}}}\right){\rm \ g\ cm\ s^{-2}}
\label{eq:momentum2}
\end{array}
\end{equation}
Therefore we argue that it is unlikely that the outflow is entirely momentum-driven. However, given the uncertainty in the mass-outflow rate (Section \ref{sec:drivingmechanism}), we cannot rule out this possibility.

Alternatively, (part of) the fast outflow in region E0 could be energy-driven by a jet-induced lobe expansion. In this scenario the gas is swept up and compressed as the radio jets hollow out a cocoon-like structure \citep[e.g.][]{cap99,tad01}. According to \citet{vil99} outflow velocities $\ga 1000$ km~s$^{-1}$ can more easily be explained if the clouds are entrained in hot, shocked gas that expands out {\sl behind} the bow-shock, similar to the expansion of an interstellar bubble \citep[see also e.g.][]{sto92,dai94,kle94,ode03}. This process has already been suggested for several other radio galaxies \citep[e.g.][]{vil99,bes00}. Such a more gentle energy-driven mechanism might also be better for explaining the large amounts of neutral gas involved in the outflow.

\subsubsection{Neutral gas involved in the outflow}

The fact that only about 1 per cent of the outflowing gas appears to be ionized, 
while the rest is in its neutral state, is maybe one of the most intriguing results from this paper 
and Paper 1. The question is how, despite the high energies that must be involved in a jet-ISM 
interaction, such large amounts of gas stay, or become again, neutral? 

The answer may lie within various simulations of jet-cloud interactions. For example, 
simulations by \citet{sut03} show that post-shock gas overtaken by a radio jet can show complex 
cooling. In the fractal structure of the gas dense filaments can be formed. 
Taking into account cooling effects, \citet{sax05} and \citet{bic03} show that in 
various jet-ISM interactions intervening clouds can severely disrupt the radio jet, while 
at the same time some clouds are accelerated away from their initial positions by the thrust 
of the radio jet. Large clouds, in particular the ones not directly in the path of the jet, 
can survive for a long time after the radio jet passed. Simulations by \citet{mel02} investigate the fate 
of a cloud of ionized gas that is overrun by the lobe of an expanding radio jet. Due to an underpressure in the cloud compared to the overpressured cocoon, shocks start to travel into the cloud and the cloud is compressed and fragmented. Due to rapid cooling the bulk of gas gets locked up in these dense fragments that can survive for a long time and reach speeds up to 500 km~s$^{-1}$. The further evolution of these fragments will be dominated by further acceleration and erosion by the passing flow, as well as gravitational collapse (which ultimately will result in star formation). \citet{fra04} reach similar conclusions from their models on jet-induced star formation. They show that in the interaction of shocks with a radiative cloud, a large fraction of the cloud-gas cools to low temperatures. The neutral hydrogen gas in the starforming Minkowski's Object, for example, may have cooled from the initial warm gas as a result of the radiative cooling triggered by a passing radio jet. The amount of outflowing neutral hydrogen gas that we find in region E0 in 3C~293 might be explained in a similar way, although in the simulations by \citet{fra04} the post-shock gas is only accelerated very slowly.

In light of the fast outflow of neutral hydrogen gas that we find in 3C~293, 
it would be interesting to explore whether, in the above mentioned simulations, at least a significant part of the dense clouds or fragments can consist of gas that remained or recombined again into neutral hydrogen gas, and if this gas can reach velocities up to 1000 km s$^{-1}$ during the lifetime of the radio source.}

\end{itemize}

\subsection{Outflows in other regions}
\label{outflowother}

As explained in Section \ref{sec:radioaxis}, there are other regions in which
complex kinematics of the gas, possibly connected to outflows, are seen. However, 
the gas kinematics observed in the other regions are not as extreme as in region E0. 

The propagating radio jet, responsible for the fast outflow in region E0 as explained above, 
likely carries enough energy to also cause outflows in the regions other than E0. This is an 
appealing scenario, in particular since the complex kinematics are observed along the inner radio axis. The fact that the broad component of the emission-line gas (the component that represents the ionized gas outflows) is blueshifted w.r.t. the narrow component on the eastern side and redshifted on the western side (in case the asymmetric emission-line profile in regions W1 and W2 represents a real outflow and is not the effect of observational line-of-sight effects as discussed in Section \ref{sec:radioaxis}), implies that the radio jet is approaching on the eastern side and receiding on the western side. However, in region E2 so far no radio continuum has been detected (see Section \ref{sec:radioaxis}), 
although it is the region in between the bright inner radio jet and the bright outer radio lobe. This could mean that the radio jet is propagating more efficiently in this region and it 
might suffer from Doppler de-boosting. Something similar likely happens in the region between the 
core and the region E0 as observed with very high resolution VLBI imaging \citep{bes04}. Another 
possibility is that the inner jet represent re-started activity \citep[as suggested by][]{aku96}, 
and the radio continuum in between the inner jet and the outer lobe has faded away. This last 
explanation could resemble the case of Centaurus~A. As in 3C~293, also in Centaurus~A a so 
called inner-filament of ionized gas with very complex kinematics \citep{mor91} is located 
outside the inner bright radio lobe (the lobe that can be considered to be the result of 
the most recent activity), but about 2 kpc from the much fainter large-scale jet \citep{mor99}. In the case of Centaurus~A it has been proposed that the large velocities shown by the gas are still 
created by the strong instabilities produced by the propagation of the jet in the past (especially in the 
transition region between the inner and the outer lobe). A similar scenario could apply to the outflow observed in region E2 in 3C~293.

Using similar arguments to those used for region E0, the outflow in region E2, although 
it appears to have broken through the ionized disk of the galaxy, is 
still not spread widely enough that it can be explained by a fossil wind from the 
post-starburst stellar population found by \citet{tad05}. Also, from both our optical 
spectra as well as the UV-data from \citet{all02}, there is no indication that a substantially 
younger stellar population or \HII\ region is present in region E2. Therefore it is not likely that the outflow in region E2 is driven by stellar winds. 

An AGN-wind is not likely strong enough to explain the outflow in region E2, since 
the outflow seems to be too extreme for that. However, for the outflow in region 
E1 and in particular regions W1 and W2 (assuming the outflow is real in these 
western regions and not a result of observational line-of-sight effects as discussed in 
Section \ref{sec:radioaxis}) we cannot discard an AGN-wind as the possible 
driving mechanism with the available data. 

\section*{Acknowledgments}

We are grateful to T. Robinson for the reduction of the p.a. 60$^{\circ}$ and 135$^{\circ}$ spectra. BE would like to thank K. Wills for her help and advice, G. Bicknell and R. Sutherland for a very useful discussion and for giving good insights into jet-ISM interactions, and M. Villar-Martin for her tips on line-ratio analysis. We would also like to thank R. Beswick for providing his nice MERLIN-HST overlay and the support astronomers at the WHT for taking the high-quality spectra during service time. BE acknowledges the University of Sheffield and ASTRON for their hospitality during this project. Part of this research was funded by the Netherlands Organization for Scientific Research (NWO) under grant R 78-379.

\end{document}